# On strong *f*-electron localization effect in topological kondo insulator

*Udai Prakash.Tyagi, Kakoli Bera, and Partha Goswami(R)\**

*D.B.College, University of Delhi, Kalkaji, New Delhi-110019, India*
*\*Email of the corresponding author: physicsgoswami@gmail.com*
*Email of first author: uptyagi@yahoo.co.in*
*Email of second author: kakolibera@gmail.com*

## Abstract

We study strong *f*-electron localization effect on surface state of a generic topological Kondo insulator (TKI) system by performing a mean-field theoretic (MFT) calculation within the frame-work of periodic Anderson model (PAM) using slave-boson technique. The surface metallicity together with bulk insulation is found to require this type of localization. A key distinction between surface states in a conventional insulator and a topological insulator is that, along a course joining two time-reversal invariant momenta (TRIM) in the same BZ, there will be intersection of these surface states, even/odd number of times, with the Fermi energy inside the spectral gap. For even (odd) number of surface state crossings, the surface states are topologically trivial (non-trivial). The symmetry consideration and the pictorial representation of surface band structure obtained here show odd number of crossing leading to the conclusion that, at least within PAM framework, the generic system is a strong topological insulator.

## 1. Introduction

We investigate the effects of strong *f*-electron localization on surface state of a generic topological Kondo insulator (GTKI). The boride $SmB_6$ is a typical example of this class of compound. Its d band metallic counterpart is $LaB_6$. In fact, it was proposed in 2008, after the discovery of topological insulators (TIs), that $SmB_6$ is a topological band insulator[1] supporting metallic surface states with an insulating bulk. The angle-resolved photoemission spectroscopy (ARPES)[2,3] and the transport measurement [4] on $SmB_6$ seem to authenticate the topological interpretation of $SmB_6$, providing persuasive affirmation of metallic surface states with an insulating bulk. In this paper, our focal point is a generic topological Kondo insulator (GTKI) and its band structure. We study strong *f*-electron localization effect on surface state of this system. Time to time, in the course of discussion, we shall refer to $SmB_6$ to clarify issues which come into view. For example, the issue of categorization into weak and strong TI. A TKI like $SmB_6$ needs to display an odd-number of band-crossings between the Sm 4f- and 5d-bands (or, is it overwhelmingly 2p B ?[5]) for being categorized as a strong TI; an even number of crossings corresponds to a weak TI. Our analysis is based on the Periodic

---------------------------------------------------------------------------------------------------

The official email id of the corresponding author: pgoswami@db.du.ac.in; The official email id of the first author: uptyagi@db.du.ac.in; The official email id of the second author: kbera@db.du.ac.in .

Anderson Model (PAM) **[6,7]** well-suited to discuss a Kondo insulator (KI)**[7]**. We shall show in section 3 that the strong *f*-electron localization makes GTKI fall decidedly into the former category.

A KI basically corresponds to periodic array (Kondo lattice) of localized spin states, which hybridize with itinerant electron sea leading to an energy gap in the electronic density of states (DOS) whose magnitude is strongly temperature-dependent and only fully developed at low temperatures. Interestingly, KIs display metallic behavior at high temperature regime. Upon reduction of temperature, the resistivity decreases up to a minimum value followed by a divergence of the form -ln(T). As had been concluded by Konndo[8] that such a strange variation can be ascribed to strong dependence on the conduction electron scattering by the localized magnetic moments. The s-d hybridization model put forward by Kondo displays a similar trend in resistivity up to a certain temperature, generally referred to as the Kondo temperature ($T_k$). At temperatures less than $T_k$, the resistivity is found to diverge. The proximity of periodic ions brings into play additional strong correlation effect (SCE) which leads to the renormalization of band structure and reconstruction of the hybridization gap in the density of states (DOS). While for heavy fermion metals the Fermi level coincides with a finite DOS(the chemical potential is located in the conduction band), for KI the Fermi level falls in the hybridization gap which could be either direct or indirect**[9]**. Most studied Kondo insulators**[9]** are FeSi, $Ce_3Bi_4Pt_3$, **$SmB_6$**, $YbB_{12}$, and CeNiSn. Upon decreasing the temperature further, there would be coupling of the spins of the itinerant electrons and those of the impurities leading to the spin-flip scattering (SFS)and the disappearance of the -ln(T) behavior. The near disappearance of local magnetic moments, too, will occur due to the screening by the sea of conduction electrons. In fact, the spin-spin coupling (and SFS) also causes formation of spin singlets. This eventually leads to typical Doniach-like phase diagram[10] (see section 2)**.** The topological Kondo insulator **[11,12]**(TKI), on the other hand, are a class of narrow gap insulator in which the gap is created by the strong *f*-electron correlations but, which are at the same time, topologically ordered due to large spin orbit coupling (and the odd(even)-parity of the localized *f*-states (conduction band)). The strongly localized *f* electrons and the spin-momentum locked helical liquid-like band structure in the surface of these systems give rise to exquisite electronic properties. In Figure 1 we have shown a qualitative sketch of DOS ($eV^{-1}$) as a function of energy (eV) for a GTKI. It must be mentioned that the hybridization gap on Sm-terminated and B-terminated surfaces exhibit very similar behavior including display of a finite DOS at E =0, compatible with an additional (surface) conductance channel.   Supposing Figure 1 corresponds to a Sm-terminated surface, all other features get shifted toward smaller absolute values of energy on a B-terminated surface.

The periodic Anderson model **[6,7]**, used in this communication for a GTKI **[1-5]**, ignores the complicated multiplet structure of the *d* and *f* orbitals usually encountered in   real TKIs, such as $SmB_6$. On a quick side note, we present the structure of   $SmB_6$ to apprise reader with complications involved:   Electrons in a rare earth occupy shells of either $[Xe]4f^n5d^16s^2$or $[Xe]4f^{(n+1)}5d^06s^2$. The states have multiplets of energy levels: Sm*f* levels split into $4f_{5/2}$ and $4f_{7/2}$ bands with large energy separation. There is strong onsite repulsion of *f* electrons. We assume in PAM that the *f* electrons locally interact via a Hubbard-*U* repulsion while the *d* electrons are practically non-interacting. The crystal field splits Sm $4f_{5/2}$ bands into a $\Gamma_7$ doublet and a $\Gamma_8$ quartet. Away from the $\Gamma$ point, the $\Gamma_8$ quartet further splits into $\Gamma_8^1$ and $\Gamma_8^2$ doublets. Furthermore, Sm $5d$ states exhibit with $t_{2g}$ and $e_g$ symmetry. These splitting are depicted in a cartoon caricature in Figure 2. In a realistic description, the starting Hamiltonian H must involve terms representing this complex multiplet structure of the $d$ and $f$ orbitals including the strong correlation effect. In that case, however, the analysis will be an unenviable task. Besides, such an analysis is beyond the scope of this communication. We note that the 4f electrons are closer to the nucleus in comparison with the 5d or 6s electrons. Thus, the electrons in the 4f shell are more localized (than the 5d or 6s electrons) and, consequently, their orbital angular momentum behaves like that of a free atom. On the other hand, the outer ones, such as the ones in 5d orbital are in the midst of the crystalline environment. For this reason, their angular momentum average is almost zero due to its precession in the crystal field. As regards the boride $SmB_6$, it possesses inversion symmetry (see section 3). The $Z_2$ topological invariants have been computed via parity analysis by previous workers[2-4] and found to be $Z_2 = 1$. On the basis of this, it was predicted by Dzero et al.[13-16] that $SmB_6$ is a topologically non-trivial system. Furthermore, within the bulk hybridization gap, signature of two dimensional Fermi surfaces on (100) and (101) surface planes supporting the presence of topological surface states were obtained in the quantum oscillation experiments of Li et al.[17]. While on quantum oscillations, as a side note, it could be mentioned that Tan et al. [18], and subsequently Sebastian et al. [19], have uncovered a deep mystery associated with the insulating bulk of $SmB_6$ which is the formation of a large three-dimensional (solely) conduction electron Fermi surface (FS), given that thus far such FSs have been considered the preserve of metals, in the absence of the long-range charge transport. This is possibly due to the residual density of states at the Fermi energy (shown in Figure 1) in $SmB_6$. The existence of the residual DOS was indicated through measurements of heat capacity[20] long ago. Now coming back to the original track, we note that recent study[5] by Maiti et al., however, argued that the observed metallic surface states have trivial origin rendering $SmB_6$ a trivial surface conductor. The symmetry consideration and the pictorial representation of surface band structure obtained show odd number of crossing (presented in section 3 of this paper) which leads to the conclusion that this is not true for a GTKI.

We now present the organization of the paper in somewhat details: In section 2, we start with dispersive f, and d bands involving hopping ($t_f$, $t_d$) for f, and d electrons with $t_d \gg t_f$. The $d$ bands corresponds to bath which can be described by Bloch states. The assumption that correlation between electrons on the impurity ion ($f$ band) much greater than all hopping terms has the effect of favoring single occupation of the impurity level. The correlation, being important for the formation of a magnetic moment on the localized $f$-orbital, arises from the Coulomb repulsion $U \gg t_d$ between the electrons. The next term involves f band, and d/p bands which hybridize at cryogenic temperatures (a strong spin-orbit overlap) forming an insulating gap at with the Fermi level residing in the hybridization gap. The hybridization V, between an odd-parity nearly localized band and an even-parity delocalized conduction band, plays the role of band-inversion yielding a 3D TI. These terms together constitute PAM Hamiltonian where there are two different species of electrons, namely conduction electrons and localized electrons. When U is very large ($U \gg t_d$) and $V \ll t_d$, the weight of configurations with the number of f electrons significantly different from its average number is very few and far between in the ground-state wave function and charge fluctuations become effectively non-existent. This results in transformation of the PAM Hamiltonian $H^{PAM}$ to Kondo lattice Hamiltonian $H_K$ [21,22] that describes interaction between spins of localized and conduction electrons. The Hamiltonian $H_K$ is obtained by using a second-order perturbation with respect to hybridization V of PAM where the localized f electrons can exchange with the conduction electrons bath thus allowing both charge and spin fluctuations to occur. Whereas the Hamiltonian H$_K$ leads to Kondo singlet formation, H$^{PAM}$ constitutes our minimalistic, starting Hamiltonian H$^{bulk}$ which captures essential physics of GTKI in the presence of strong *f*-electron localization, such as the bulk and surface band-structure topology. In section 3 we investigate the surface state Hamiltonian $H_s$ in slab geometry starting from H$^{bulk}$. We use the Dirac-matrix Hamiltonian (DMH) method [23] to ascertain whether SmB$_6$ is a weak or strong TKI. It must be emphasized here that when a slab of finite thickness is considered, two surface states overlap such that off-diagonal mass-like term $M_0$ must be included in the surface Hamiltonian $H_s$ above. For this the matrix $\gamma_5 M_0$ is suitable, where $\gamma_5 \equiv i\gamma_0\gamma_1\gamma_2\gamma_3$ and $\gamma_i$ are Dirac matrices in Dirac basis. As an important step of the DMH method, one needs to check anticommutativity of the matrix $\gamma_5 M_0$ with $H_s$. We obtain here $\{H_s, \gamma_5 M_0\} \neq 0$, and therefore the topological stability of the band crossing is possible [23]. The combined surface state Hamiltonian that we proceed with is $H^{slab}(k_x, k_y) = H_s(k_x, k_y) + \gamma_5 M_0$. We find that $H^{slab}(k_x, k_y)$ preserves time reversal symmetry (TRS) and the inversion symmetry (IS). The pictorial representation of surface band structure obtained by us leads to the conclusion that, at least within single impurity PAM framework, GTKI system under consideration is a strong topological insulator. The paper ends with discussion and concluding remarks in section 4.

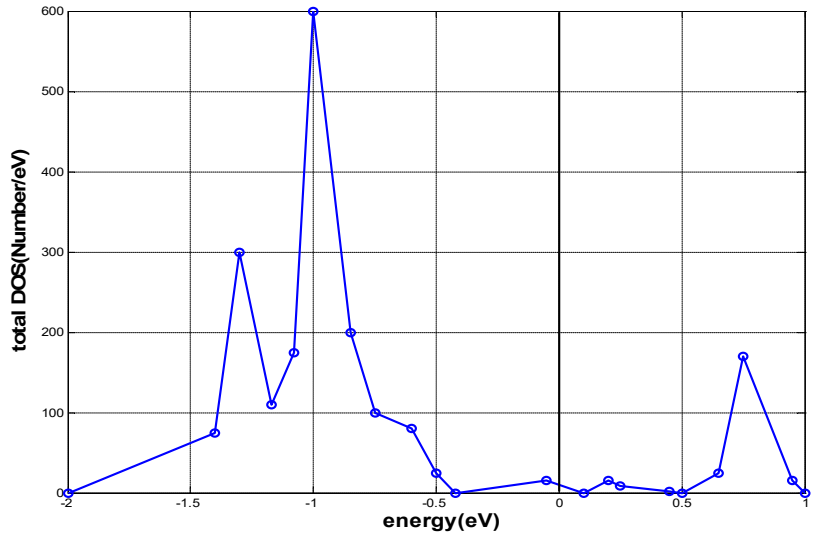

**Figure 1.** A qualitative sketch of DOS (eV$^{-1}$) as a function of energy (eV) for a GTKI. While for TI and TKI the Fermi level falls in the hybridization gap of total density of states (DOS), heavy fermion metals the Fermi level coincides with a finite DOS. The position of the Fermi energy $E_F$ is set at zero energy. There is a finite DOS-value at energy $E = E_F$. This might indicate an incomplete (or pseudo-) gap, but is also compatible with an additional conductance channel at the surface.

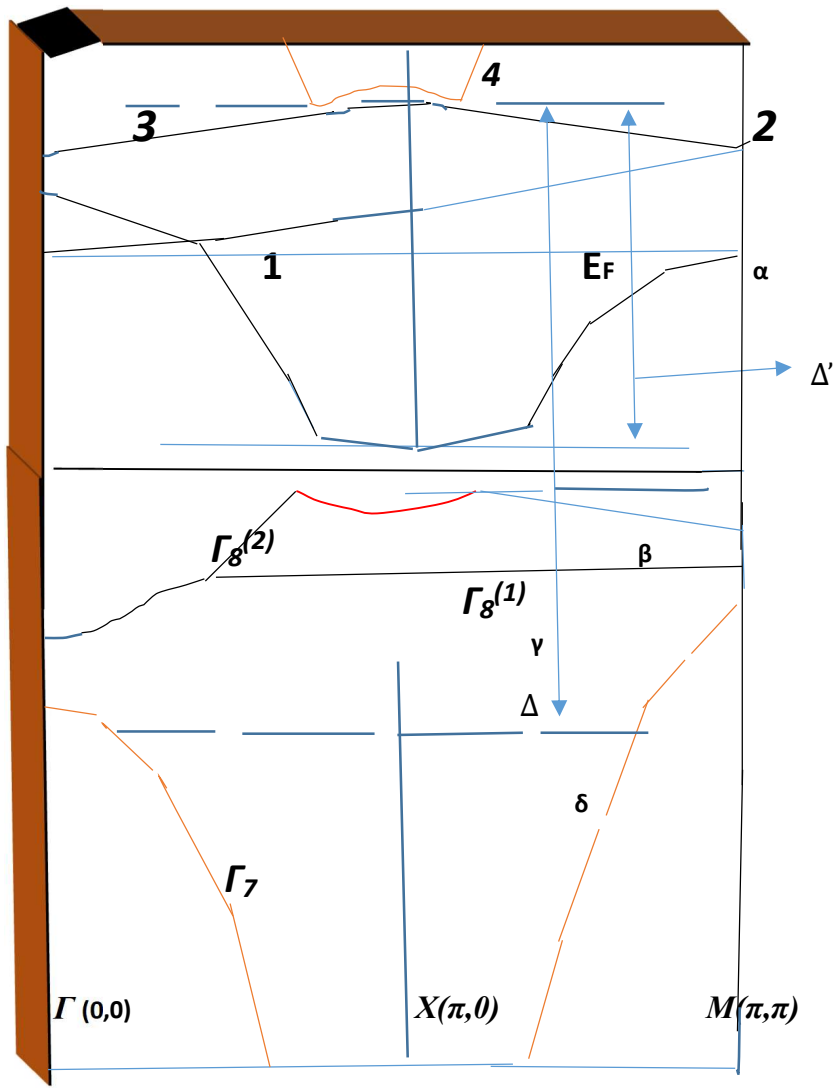

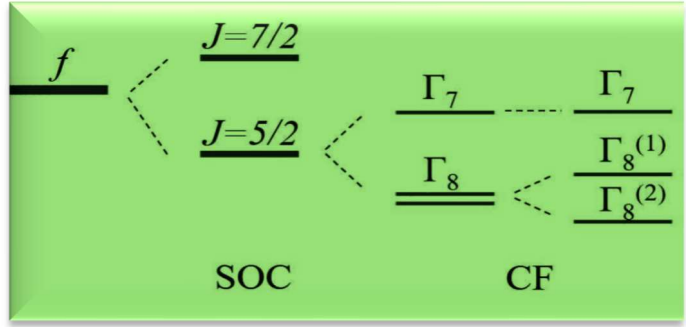

**Figure 2:** A cartoon caricature representation of the bulk band structure without self-energy broadening in SmB$_6$. An alternative, conventional representation of the evolution of energy levels of the *f*-states in SmB$_6$, due to intervention of the spin-orbit coupling and the crystal field, is shown in the lower panel for greater clarity. The *f*-states are split into $J = 7/2$ and $J = 5/2$ states by spin-orbit coupling (SOC). The $J = 5/2$ state, slightly below the Fermi energy $E_F$, is split into a $\Gamma_7$ doublet and a $\Gamma_8$ quartet by the crystal field (CF). Away from the $\Gamma$ point, the $\Gamma_8$ quartet is further split into $\Gamma_8^{(1)}$ and $\Gamma_8^{(2)}$ doublets. The hybridization between the $\Gamma_7$, $\Gamma_8^{(1)}$ bands and the conduction band opens two gaps which are denoted as $\Delta$ (typically around 15 meV) and $\Delta' \sim 3 meV$.

## 2. Kondo Screening

A Kondo system exhibits the Kondo screening. So, while investigating a Kondo system the possibility of the Kondo screening needs to be explored. In an effort to do this, we calculate the Kondo singlet density below within PAM-Kondo framework. The PAM model [6,7,24] on a simple cubic lattice for *d* and *f* electrons in momentum-space is as follows:

$$\aleph = \sum_{k,\zeta\,=\,\uparrow,\downarrow}(-\mu - \epsilon_k^f)f_{k,\zeta}^\dagger f_{k,\zeta} + \sum_{k\zeta\,=\,\uparrow,\downarrow}(-\mu - \epsilon_k^d)d_{k,\zeta}^\dagger d_{k,\zeta}$$
$$+ \sum_{k,\zeta=\uparrow,\downarrow}\{\Gamma_{\zeta\,=\,\uparrow,\downarrow}(\boldsymbol{k})\,d_{k\zeta}^\dagger f_{k,\zeta} + \text{H.C.}\} + \aleph_{int} \qquad (1)$$
$$\epsilon_k^d = [2t_{d1}\,c_1(k) + 4t_{d2}\,c_2(k) + 8t_{d3}\,c_3(k)],$$
$$\epsilon_k^f = [-\epsilon_f + 2t_{f1}\,c_1(k) + 4t_{f2}\,c_2(k) + 8t_{f3}\,c_3(k)],$$

$$c_1(k) = (\cos k_x a + \cos k_y a + \cos k_z a),$$
$$c_2(k) = (\cos k_x a \cos k_y a + \cos k_y a \cos k_z a + \cos k_z a \cos k_x a),$$
$$c_3(k) = (\cos k_x a \cos k_y a \cos k_z a),$$
$$\aleph_{int} = U_f \sum_{i(site\ index)} f^\dagger_{i\uparrow} f_{i\uparrow} f^\dagger_{i\downarrow} f_{i\downarrow}\ . \qquad (2)$$

We have assumed that, whereas the coulomb repulsion between *f* electrons on the same site could be approximated by a term involving the Hubbard-*U*, the d electrons are practically non-interacting. In rest of the terms, in momentum-space, *d* and *f* electrons are represented by creation (annihilation) operators $d^\dagger_{k\zeta}$ ($d_{k\zeta}$) and $f^\dagger_{k\zeta}$ ($f_{k\zeta}$), respectively. Here, the index $\zeta$ (= ↑,↓) represents the spin or pseudo-spin of the electrons with *a* as the lattice constant. The first and the second term, respectively, describe the dispersion of the *f* and *d* electrons. The parameters ($t_{f1}$, $t_{f2}, t_{f3}$), and ($t_{d1}, t_{d2}, t_{d3}$) correspond to the (*NN, NNN, NNNN*) hopping for the *f* and *d* electrons, respectively. The hybridization between the *f*-electrons (*l* =3 and hence odd parity) and the conduction d-electrons (*l* =2 and hence even parity) **[6,7,24]** is given by the third term. The parity is a good quantum number. It follows that at high symmetry points (HSP) in the Brillouin zone (BZ), where odd and even parity states cannot mix, the third term is zero. The presence of the hybridization node at HSP is an important feature of TKI. Furthermore, since the *f*- and *d*-states have different parities, the momentum-dependent form-factor matrix $\Gamma_{\zeta=\uparrow,\downarrow}(k)$ involved in the third term in (1) must be odd: $\Gamma_{\zeta=\uparrow,\downarrow}(-\boldsymbol{k}) = -\Gamma_{\zeta=\uparrow,\downarrow}(\boldsymbol{k})$. This is very much required in order to preserve time reversal symmetry (TRS), as the matrix involves coupling with the physical spin of the electron. Therefore, we write $\Gamma_{\zeta=\uparrow,\downarrow}(k) = -2V(\boldsymbol{s}(k).\boldsymbol{\zeta})$, where V is a constant parameter characterizing the hybridization, $\boldsymbol{s}(k) = (\sin k_x a, \sin k_y a, \sin k_z a)$ and $\boldsymbol{\zeta} = (\zeta_x, \zeta_y, \zeta_z)$ are the Pauli matrices in physical spin space. In section 4, we have outlined what to use as more complicated odd-parity expressions for $\Gamma$, to re-investigate the present problem, expecting newer information.

The system shows the bulk metallic as well as the bulk insulating phases determined by the sign of $t_{f1}$. It is positive for the former and negative for the latter phase **[24]**. The negative sign of $t_{f1}$ is also necessary for the band inversion, which induces the topological state **[25]**. Since the bandwidth of the *f* electrons needs to be smaller than the bandwidth of the conduction electrons, we assume that $|t_{f1}| << |t_{d1}|$. Similar relations hold for second- and third-neighbor hopping amplitudes. For the hybridization parameter *V* we assume $|V| < |t_{d1}|$. **Throughout the whole paper, we choose $t_{d1}$ to be the unit of energy, except in the investigation of Kondo screening.** We further assume that, while the *d* electrons are non-interacting, the on-site coulomb repulsion of *f* - electrons is given by the Hubbard U **[24]**. Under the assumption that U is considerably greater than $|t_{d1}|$, one may write the thermal average of the TKI slave-boson mean-field Hamiltonian **[24]** in the form (see also section 4)

$$\langle \aleph_{sb}(b,\lambda,\xi) \rangle = \sum_{k\zeta\ =\uparrow,\downarrow} (-\mu - \xi - \epsilon^d_{\boldsymbol{k}}) \langle d^\dagger_{k,\zeta} d_{k,\zeta} \rangle$$
$$+ \sum_{k,\zeta\ =\uparrow,\downarrow} (-\mu + \xi - b^2 \epsilon^f_{\boldsymbol{k}} + \lambda) \langle s^\dagger_{k,\zeta} s_{k,\zeta} \rangle$$

$$+ b\sum_{k,\zeta,\sigma\ =\uparrow,\downarrow} \{\Gamma_{\zeta=\uparrow,\downarrow}(\boldsymbol{k})\ \langle d^\dagger_{k\zeta} s_{k,\zeta} \rangle + \text{H.C.}\} + \lambda N_s (b^2 - 1) \qquad (3)$$

where the additional terms, in comparison with (1), are $\lambda \, [\sum_{k,\sigma \,=\, \uparrow,\downarrow} \langle s^\dagger_{k,\zeta} s_{k,\zeta}\rangle + N_s (b^2 - 1)] - \xi [N_d - N_s] - \mu [(N_d + N_s) - N]$ where $\lambda$ is a Lagrange multiplier, $\mu$ is the chemical potential, $N_d$ is the number of lattice sites for $d$ electrons, and $N_s$ corresponds to that for $f$-electrons. The first term describes the constraint on the pseudo-particles due to the infinite Coulomb repulsion. This ensures strong localization of $f$ electrons. The second term compels observance of the fact that the $d$ and $f$ fermions are equal in number ($N_d = N_s$). The reason simply is the key requirement for a Kondo insulator, viz. the formation of singlet pairing states between $d$ and $f$ fermions at each lattice site for which the number of $d$ and $f$ fermions must be equal on average. The third term is the constraint which fixes the total number of particles $N$ ($N = N_d + N_s$). An explanation to highlight the physical mechanism underlying the second constraint, and the physical interaction that enforces it, perhaps will not be irrelevant: We are investigating the Kondo screening under the assumption "large on-site repulsion ($U_f \gg t_{d1}$) between the $f$-electrons" (and no interaction $U_d$ between conduction electrons). In this situation each of the impurities simply becomes an isolated magnetic moment correlated with an itinerant electron spin. The necessity of "chemical potential" $\xi$ is rooted in the correlation between impurity and conduction electron spins. The interaction involved is the spin-spin exchange interaction. It is clear from Eq.(3) that the hybridization parameter is renormalized by the $c$-number $b$. It is also clear that the dispersion of the $f$-electron is renormalized by $\lambda$ and its hopping amplitude by $b^2$. The total parameters are, thus, the Lagrange multiplier $\lambda$, auxiliary chemical potentials $\xi$ and $\mu$ ($\mu$ is a free parameter), and slave-boson field $b$. In order to make this paper self-contained we found necessary to include these facts, though these are explained clearly in ref. **[24]**. The minimization of the thermodynamic potential per unit volume with relative to the parameters ($b, \lambda, \xi$) yields equations to determine them: $\Omega_{sb} = -(\beta V)^{-1} \ln Tr \, exp \, (-\beta \aleph_{sb}(b, \lambda, \xi))$; $\partial \Omega_{sb}/\partial b = 0, \partial \Omega_{sb}/\partial \lambda = 0$, and $\partial \Omega_{sb}/\partial \xi = 0$, where $\beta = (k_B T)^{-1}$, $k_B$ is Boltzmann constant and T is temperature. The method outlined in refs.**[26,27]** has been used to calculate the thermodynamic potential.

The equations $\partial \Omega_{sb}/\partial b = 0$, $\partial \Omega_{sb}/\partial \lambda = 0$, and $\partial \Omega_{sb}/\partial \xi = 0$, respectively, can be written as $2\lambda b = N_s^{-1}(\alpha_1 + \alpha_2 + \alpha_3)$, $(1 - b^2) = N_s^{-1}(\beta_1 + \beta_2 + \beta_3)$, and $0 = N_s^{-1}(\gamma_1 + \gamma_2 + \gamma_3)$ where

$$\alpha_1 = \sum_k \frac{\partial}{\partial b}[2V(s_x - i\,s_y)\langle d^\dagger_{k\uparrow} bs_{k,\downarrow}\rangle + 2V(s_x - i\,s_y)\langle bs^\dagger_{k,\uparrow} d_{k\downarrow}\rangle + H.C.]$$

$$\alpha_2 = \sum_{k,\zeta} \frac{\partial}{\partial b}[\epsilon^f_k \langle bs^\dagger_{k,\zeta} bs_{k,\zeta}\rangle + \epsilon^d_k \langle d^\dagger_{k,\zeta} d_{k,\zeta}\rangle]$$

$$\alpha_3 = \sum_{k,\zeta} \frac{\partial}{\partial b}[2V s_z \zeta \langle d^\dagger_{k,\zeta} bs_{k,\zeta}\rangle + H.C.], \qquad (4)$$

$$\beta_1 = \sum_k \frac{\partial}{\partial \lambda}[2V(s_x - i\,s_y)\langle d^\dagger_{k\uparrow} bs_{k,\downarrow}\rangle + 2V(s_x - i\,s_y)\langle bs^\dagger_{k,\uparrow} d_{k\downarrow}\rangle + H.C.]$$

$$\beta_2 = \sum_{k,\zeta} \frac{\partial}{\partial \lambda}[\epsilon^f_k \langle bs^\dagger_{k,\zeta} bs_{k,\zeta}\rangle + \epsilon^d_k \langle d^\dagger_{k,\zeta} d_{k,\zeta}\rangle]$$

$$\beta_3 = \sum_{k,\zeta} \frac{\partial}{\partial \lambda}[2V s_z \zeta \langle d^\dagger_{k,\zeta} bs_{k,\zeta}\rangle + H.C.], \qquad (5)$$

$$\gamma_1 = \sum_k \frac{\partial}{\partial \xi}[2V(s_x - i\,s_y)\langle d^\dagger_{k\uparrow} bs_{k,\downarrow}\rangle + 2V(s_x - i\,s_y)\langle bs^\dagger_{k,\uparrow} d_{k\downarrow}\rangle + H.C.]$$

$$\gamma_2 = \sum_{k,\zeta} \frac{\partial}{\partial \xi} [\epsilon_k^f \langle bs_{k,\zeta}^\dagger bs_{k,\zeta}\rangle + \epsilon_k^d \langle d_{k,\zeta}^\dagger d_{k,\zeta}\rangle$$

$$\gamma_3 = \sum_{k,\zeta} \frac{\partial}{\partial \xi} [2Vs_z\zeta \langle d_{k,\zeta}^\dagger bs_{k,\zeta}\rangle + \text{H.C.}] \tag{6}$$

$$\epsilon_k^d = [2t_{d1} c_1(k) + 4t_{d2} c_2(k) + 8t_{d3} c_3(k)] \tag{7}$$

$$\epsilon_k^f = [-\epsilon_f + 2t_{f1} c_1(k) + 4t_{f2} c_2(k) + 8t_{f3} c_3(k)]. \tag{8}$$

The averages $\langle d_{k,\zeta}^\dagger d_{k,\zeta}\rangle$, $\langle bs_{k,\zeta}^\dagger bs_{k,\zeta}\rangle$, etc. have been calculated below in the finite-temperature formalism. Here the time evolution an operator O is given by O(τ)=exp($\aleph$τ) O exp($-\aleph$τ) where τ is imaginary time. The equations for the operators { $d_{k,\zeta}(\tau), s_{k,\zeta}(\tau)$} can be written down easily, for the Hamiltonian is completely diagonal. Starting with this Hamilton, the thermal averages in the equations above are determined in a self-consistent manner. The Green's functions $G_{sb}(k\zeta, k\zeta, \tau) = -\langle T_\tau\{d_{k\zeta}(\tau)d^\dagger_{k\zeta}(0)\}\rangle$, $F_{sb}(k\zeta, k\zeta, \tau) = -b^2\langle T_\tau\{s_{k,\zeta}(\tau)d^\dagger_{k,\zeta}(0)\}\rangle$, etc., are of primary interest. Here $T_\tau$ is the time-ordering operator acting on imaginary times τ. It arranges other operators like $d_{k,\zeta}(\tau)$, *etc.* from left to right with descending imaginary time arguments. We obtain

$$G_{sb}(k\uparrow, k\uparrow, \tau \to 0^+) = u_{k,+}^{(-)^2} f_-^{(-)}(k,\mu) + u_{k,-}^{(-)^2} f_+^{(-)}(k,\mu),$$

$$G_{sb}(k\downarrow, k\downarrow, \tau \to 0^+) = u_{k,+}^{(+)^2} f_-^{(+)}(k,\mu) + u_{k,-}^{(+)^2} f_+^{(+)}(k,\mu),$$

$$F_{sb}(k\uparrow, k\uparrow, \tau \to 0^+) = (u_{k,+}^{(+)^2} - v_k^{(+)^2})f_-^{(+)}(k,\mu) + (u_{k,-}^{(+)^2} + v_k^{(-)^2}) f_+^{(+)}(k,\mu)$$

$$+ v_k^{(+)^2} f_+^{(-)}(k,\mu) - v_k^{(-)^2} f_-^{(-)}(k,\mu),$$

$$F_{sb}(k\downarrow, k\downarrow, \tau \to 0^+) = (u_{k,-}^{(+)^2} + v_k^{(+)^2})f_+^{(-)}(k,\mu) + (u_{k,+}^{(+)^2} - v_k^{(-)^2})f_-^{(-)}(k,\mu)$$

$$+ v_k^{(-)^2} f_+^{(+)}(k,\mu) - v_k^{(+)^2} f_-^{(+)}(k,\mu),$$

$$f_-^{(-)}(k,\mu) = (e^{\beta(\epsilon_-^{(-)}(k)-\mu)} + 1)^{-1}, f_+^{(-)}(k,\mu) = (e^{\beta(\epsilon_+^{(-)}(k)-\mu)} + 1)^{-1}$$

$$f_-^{(+)}(k,\mu) = (e^{\beta(\epsilon_-^{(+)}(k)-\mu)} + 1)^{-1}, f_+^{(+)}(k,\mu) = (e^{\beta(\epsilon_+^{(+)}(k)-\mu)} + 1)^{-1}. \tag{9}$$

Here

$$u_{k,\pm}^{(\zeta)^2} = \frac{1}{2}[1 \pm \frac{(2\xi+\epsilon_k^d - b^2\epsilon_k^f+\lambda)}{2\{\varepsilon(k,b,\lambda,\xi)\}}], \quad v_k^{(\zeta)^2} = \frac{-2V^2 b^2 s_z^2}{\varepsilon(k,b,\lambda,\xi)[\frac{(2\xi+\epsilon_k^d-b^2\epsilon_k^f+\lambda)}{2}+\zeta \varepsilon(k,b,\lambda,\xi)]},$$

$$\varepsilon(k,b,\lambda,\xi) = \sqrt{\frac{(2\xi+\epsilon_k^d-b^2\epsilon_k^f+\lambda)^2}{4} + 4V^2 b^2(s_x^2+s_y^2+s_z^2)},$$

$$\epsilon_\alpha^{(\zeta)}(k) = -\frac{(\epsilon_k^d+b^2\epsilon_k^f-\lambda)}{2} + \alpha\sqrt{\frac{(2\xi+\epsilon_k^d-b^2\epsilon_k^f+\lambda)^2}{4} + 4V^2 b^2(s_x^2+s_y^2+s_z^2)},$$

$$\langle d^\dagger_{k\uparrow} bs_{k,\downarrow}\rangle = \frac{V(s_x+i\,s_y)}{\varepsilon(k,b,\lambda,\xi)}\left[f^{(-)}_-(k,\mu) - f^{(-)}_+(k,\mu)\right],$$
$$\langle bs^\dagger_{k,\uparrow} d_{k\downarrow}\rangle = \frac{V(s_x+i\,s_y)}{\varepsilon(k,b,\lambda,\xi)}\left[f^{(+)}_-(k,\mu) - f^{(+)}_+(k,\mu)\right], \qquad (10)$$

and $\alpha = 1$ $(-1)$ for upper band (lower band), $\zeta= \pm 1$ labels the eigenstates ($\uparrow,\downarrow$) of $\zeta_z$. The expressions for the averages $\langle d^\dagger_{k\uparrow} bs_{k,\downarrow}\rangle$ and $\langle bs^\dagger_{k,\uparrow} d_{k\downarrow}\rangle$, etc. show that in the zero-temperature and the long-wavelength limits, their contribution to the derivatives in (4), (5), and (6) are insignificant in comparison with those of $\sum_{k,\zeta}[\epsilon^f_k\langle bs^\dagger_{k.\zeta} bs_{k,\zeta}\rangle + \epsilon^d_k\langle d^\dagger_{k.\zeta} d_{k,\zeta}\rangle]$. The observation allows us to approximate the equations $\partial'\Omega_{sb}/\partial b = 0$, $\partial'\Omega_{sb}/\partial \lambda = 0$, and $\partial'\Omega_{sb}/\partial \xi = 0$ as

$$2\lambda b \approx N_s^{-1} \sum_{k,\zeta}\frac{\partial}{\partial b}[\epsilon^f_k\langle bs^\dagger_{k.\zeta} bs_{k,\zeta}\rangle + \epsilon^d_k\langle d^\dagger_{k.\zeta} d_{k,\zeta}\rangle], \qquad (11)$$

$$(1-b^2) \approx N_s^{-1} \sum_{k,\zeta}\frac{\partial}{\partial \lambda}[\epsilon^f_k\langle bs^\dagger_{k.\zeta} bs_{k,\zeta}\rangle + \epsilon^d_k\langle d^\dagger_{k.\zeta} d_{k,\zeta}\rangle], \qquad (12)$$

$$0 \approx N_s^{-1} \sum_{k,\zeta}\frac{\partial}{\partial \xi}[\epsilon^f_k\langle bs^\dagger_{k.\zeta} bs_{k,\zeta}\rangle + \epsilon^d_k\langle d^\dagger_{k.\zeta} d_{k,\zeta}\rangle] \qquad (13)$$

in these limits. The restriction $\int dr \sum_\zeta \langle s^\dagger_\zeta(r) s_\zeta(r)\rangle \cong N_s(1-b^2)$ needs to be imposed for the conservation of auxiliary particle number. This has been noted above. Here $s_\zeta(r) = N_s^{-\frac{1}{2}} \sum_k e^{ik.r} s_{k,\zeta}$. This is the fourth equation with (11)-(13) as the first three and we have four unknowns, viz. ($b, \lambda, \xi, ak_F$) where $a \approx 4.13$ Å is the lattice constant and $ak_F$ is the Fermi wave number. We shortly explain why a single value to the Fermi wave vector ($ak_F$) is assigned. In view of the fact that the chemical potential is a free parameter and could be somewhere between the valence and the conduction bands ($\mu > \epsilon^{(+)}_-(k), \epsilon^{(-)}_-(k)$), once again it is easy to see that in the long wavelength (i.e. low-lying states) and the zero-temperature limit one may write the imposed restriction as $2 N_s^{-1} \sum_{k,\zeta} 1 = b^2 - b^4$. The integration on the left-hand-side is non-trivial, for assigning a single value to the Fermi wave vector ($ak_F$) of an anisotropic (lattice) band structure whose Fermi surface(FS) may not be a sphere is inappropriate, However, as mentioned in section I, the formation of a large three-dimensional (solely) conduction electron FS takes place in the insulating bulk of SmB$_6$. Deriving support of this mysterious finding, we assume FS to be a sphere in the first approximation. This leads to the equation

$$b^2 = \tfrac{1}{2}[1\pm\sqrt{1-\left(\tfrac{8}{3}\right)(a^3 k_F^3)}]. \qquad (14)$$

After a little algebra, we find that whereas (11) and (12) together yield $\lambda = -6t_{f1} + 6b^2 t_{f1}$, Eq.(13) yields $\xi = -3t_{d1} + 3t_{f1}$. This equation for $\xi$ will be extremely useful in section (3). We now estimate ($ak_F$) in the following manner: Since the effective mass $m^*$ (Fermi velocity $v^*_F$) of the particles in low-lying states is known to be about 100m$_e$ ( < 0.3 eV-Å) **[24],** the relation $m^* = \frac{\hbar k_F}{v^*_F}$ yields ($ak_F$)~ 0.2 which is consistent with the long wavelength limit we have taken. The two values of $b^2$ obtained from (14) are 0.9430 and 0.0565. As we see below in Eq. (15), when b is nonzero the system corresponds to a Kondo state, and, when b is zero, the system is a non-interacting, dispersive lattice gas mixture of itinerant electrons and charge carrying heavy bosons( see section 4). The admissible value of $b^2$ will be thus 0.9430. As all the unknown parameters have been determined now, the stage is set to consider the Kondo screening.

We calculate the Kondo singlet density which is defined as $K_{singlet}(k,b,\lambda,\mu,\xi) = [\langle d^\dagger_{k\uparrow} bs_{k,\downarrow}\rangle + \langle bs^\dagger_{k,\uparrow} d_{k\downarrow}\rangle]$. This average is the ultimate signature of the Kondo insulating state, where there is precisely one conduction electron paired with an impurity spin. The important point now is that a large U puts down charge fluctuations on the local moment site. The corresponding "charge degree of freedom" is "quenched". The degree of freedom that remains at the impurity is its spin. Naturally, the Hamiltonian should contain a term $H_k$ comprising of the spin operator of the impurity and the conduction electron spin. Suppose, $S_m$ denotes the $m$ th-site impurity spin operator, and $s_m = \left(\frac{1}{2}\right) d^\dagger_{m\zeta}\zeta_z d_{m\zeta}$ is that for the conduction electron spin, where $d_{m\zeta}$ is the fermion annihilation operator at site-$m$ and spin-state $\zeta (=\uparrow,\downarrow)$ and $\zeta_z$ is the z-component of the Pauli matrices. The required term $H_k$ can now be written[28] as $[(-|J|/t_{d1})\sum_m S_m.s_m]$. This is an anti-ferromagnetic exchange-coupling term representing the exchange interaction between the itinerant (conduction) electrons and impurity magnetic moment in the system. For $|S| > 1$, we may approximate the impurity spins as classical vectors. This allows us to replace the exchange coupling constant $J$ by $M = -(|J||S|/t_{d1})$. It follows that the exchange field term gives the dimensionless contribution $[M \sum_{k,\zeta} sgn(\zeta) d^\dagger_{k,\zeta} d_{k,\zeta}]$ to the Hamiltonian in (3) in the basis $(d_{k\uparrow}\ f_{k\uparrow}\ d_{k\downarrow}\ f_{k\downarrow})^T$ in the momentum space. We notice that since a band electron hops on to the impurity site to gain kinetic energy or the impurity electron hops on to the band to lose kinetic energy, the spin flip during such hopping gives rise to an anti-ferromagnetic (AFM) exchange interaction term. Another important point is that in the local moment limit, the Anderson and Kondo couplings describe the same physics. A fundamental difference between them, however, is that the Anderson model includes charge fluctuations which determine the coupling, while there is absence of the spin-orbit coupling ($V \ll t_{d1}$) in the Kondo model which includes only spin-spin interactions [28]. The interaction is non-zero only in the local moment regime. The complete derivation of AFM interaction can be found in terms of the Schrieffer-Wolff transformations [28]. The averages $\langle d^\dagger_{k,\zeta} d_{k,\zeta}\rangle, \langle bs^\dagger_{k,\zeta} bs_{k,\zeta}\rangle$, etc. have been calculated below in the finite-temperature formalism as outlined above in page 8. We find

$$K_{singlet}(k,\mu) = A_- [f_-^{(-)}(k,\mu) - f_+^{(-)}(k,\mu)] + A_+ [f_-^{(+)}(k,\mu) - f_+^{(+)}(k,\mu)], \quad (15)$$

$$A_- = \frac{2V^2(s_x^2 + s_y^2)}{\varepsilon_-(k,b,\lambda,\mu,\xi)}, A_+ = \frac{2V^2(s_x^2 + s_y^2)}{\varepsilon_+(k,b,\lambda,\mu,\xi)}, \quad (16)$$

$$\varepsilon_\mp(k,b,\lambda,\xi) = \sqrt{\frac{(2\xi + \epsilon_k^d - b^2\epsilon_k^f + \lambda \mp M)^2}{4} + 4V^2 b^2(s_x^2 + s_y^2 + s_z^2)} \quad (17)$$

$$\epsilon_\alpha^{(\zeta)}(k) = -\frac{(\epsilon_k^d + b^2\epsilon_k^f - \lambda + \zeta M)}{2} +$$

$$\alpha \sqrt{\frac{(2\xi + \epsilon_k^d - b^2\epsilon_k^f + \lambda + \zeta M)^2}{4} + 4V^2 b^2(s_x^2 + s_y^2 + s_z^2)} \quad (18)$$

With the help of Eqs. (16)-(18) we obtain non-zero values of K$_{singlet}$ defined in Eq.(15). As a function of anti-ferromagnetic exchange field energy (M) and (Boltzmann constant. Temperature)( kT) in eV, we have contour/3D plotted K$_{singlet}$ in Figure 3 for (a) $\mu$ = 0.00 eV, $t_{d1}$ = 0.38 eV, and $t_{f1}$ = $-$ 0.02 eV, and (b) $\mu$ = 0.00 eV , $t_{d1}$ = 0.60 eV and $t_{f1}$= $-$ 0.03eV, at $ak_x$= $ak_y$ = $ak_z$ = 1. The anti-ferromagnetic quantum critical point (AFM QCP) is, respectively, at $M = M_C$ = 0.035 and 0.060 in (a), and (b), respectively. The two diagrams below show that location of QCP is very strongly dependent on $t_{d1}$ and $t_{f1}$. The anti-ferromagnetic quantum critical point (AFM QCP) at T = 0 K, in the former case, is at $M_{c1}$ = 0.03 eV, and in the latter case is at $M_{c2}$ = 0.060eV. The contour plot appears as Doniach-like phase diagram[10]: Close to 0 K, approximately, the region $M > M_c$ corresponds to the heavy-fermion liquid, while the region $M < M_c$ corresponds to anti-ferromagnetic liquid with quantum critical region in between. We also notice that the location of QCP depends on $\mu$ and $t_{f1}$. For example, QCP value increases with decrease in $|t_{f1}|$ when $t_{d1}$ and $\mu$ are held fixed.

## 3. Surface State Hamiltonian

The next important issue is how to get surface state Hamiltonian $\hat{H} = \aleph^{slab}(k_x, k_y)$ for slab geometry from that of bulk $\aleph^{bulk}(k_x, k_y, k_z)$ in the non-local moment limit. To explain, we first consider a slab with length, breadth, and thickness in the x, y, and z directions, respectively. The thickness is limited in $z \in [-d/2, d/2]$. We further assume the non-open boundary conditions. We can thus replace $k_z$ by $-i\partial_z$, as in this case $k_z$ is not a good quantum number. The Hamiltonian $\aleph^{slab}(k_x, k_y)$ can be obtained from the Hamiltonian $\aleph^{bulk}(k_x, k_y, k_z = -i\partial_z)$ considering a set of ortho-normal basis states $\{|\varphi_n\rangle, n = 1,2,3,4\}$ satisfying the completeness condition I= $\sum_n |\varphi_n\rangle\langle\varphi_n|$. The issue of obtaining surface state Hamiltonian is thus settled. In terms of these basis states, an arbitrary state vector $|\Phi\rangle$ can be written as $|\Phi\rangle = \sum_n |\varphi_n\rangle\langle\varphi_n|\Phi\rangle$. We now turn to the operator equation $\hat{H}|\psi\rangle = |\Phi\rangle$ where $|\psi\rangle$ an arbitrary state vector. Again, in view of the completeness condition, one can write

$$|\Phi\rangle = \hat{H} \ |\psi\rangle = \hat{H} \sum_n |\varphi_n\rangle\langle\varphi_n|\psi\rangle = \sum_n \hat{H} |\varphi_n\rangle\langle\varphi_n|\psi\rangle. \quad (19)$$

Now, form the inner product $\langle\psi|\Phi\rangle$, where $\langle\psi|\Phi\rangle = (\langle\Phi|\psi\rangle)^*$. We have

$$\langle\psi|\Phi\rangle = \sum_{m,n} \langle\psi|\varphi_m\rangle \ \langle\varphi_m|\hat{H}|\varphi_n\rangle\langle\varphi_n|\psi\rangle. \quad (20)$$

We can write the last equation as $\langle\psi|\Phi\rangle = \sum_n \psi^*_m H_{mn}\psi_n$, where $H_{mn} = \langle\varphi_m|\hat{H}|\varphi_n\rangle$ and $\psi_n = \langle\varphi_n|\psi\rangle$. Here, $\psi_n$ is an inner product and $H_{mn}$ is the Hamiltonian matrix sought for. We choose the product as

$$\left(\frac{2}{d}\right)^{\frac{1}{2}} \left[\sin\left\{\frac{n\pi\left(z+\frac{d}{2}\right)}{d}\right\}\right] \quad (n = 1, 2, 3, 4 ). \quad (21)$$

With this ansatz, the matrix elements may be written as

$$\aleph_{mn}^{slab}(k_x, k_y) =$$
$$\left(\frac{2}{d}\right)\int_{-d/2}^{d/2} dz \left[\sin\left\{\frac{m\pi\left(z+\frac{d}{2}\right)}{d}\right\}\right] \aleph_{mn}^{bulk}(k_x, k_y, -i\partial_z) \left[\sin\left\{\frac{n\pi\left(z+\frac{d}{2}\right)}{d}\right\}\right].$$
$$(22)$$

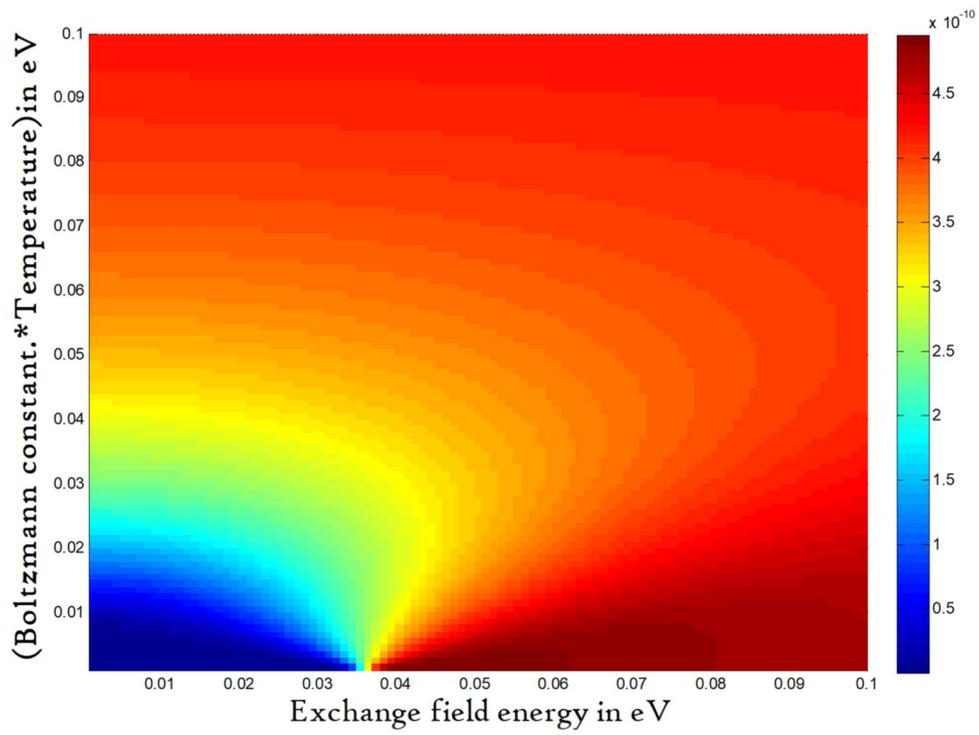

**(a)**

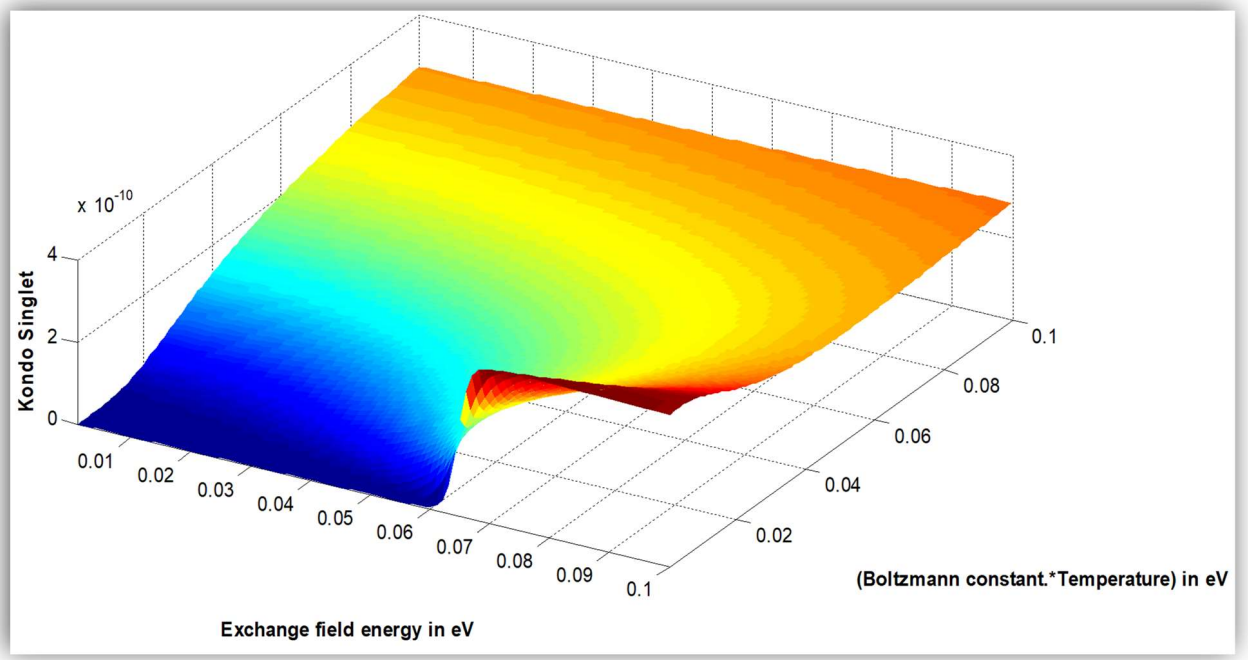

**(b)**

**Figure 3.** The contour/3D plots of Kondo singlet term given by Eq.(15) as a function of anti-ferromagnetic exchange field energy ($M$) and (Boltzmann constant. Temperature)($kT$) in eV for **(a)** $\mu$ = 0.00 eV, $t_{d1}$ = 0.38 eV, and $t_{f1}$ = − 0.02 eV, and **(b)** $\mu$ = 0.00 eV, $t_{d1}$ = 0.60 eV and $t_{f1}$ = − 0.03eV, at $ak_x = ak_y = ak_z = 1$. The anti-ferromagnetic quantum critical point (AFM QCP) is, respectively, at $M = M_C$ = 0.035 and 0.060 eV in (a), and (b). Other parameters in terms of electron-volts, in the graphical representations, are $b$ = 0.95, $t_{d2}$= 0.001, $t_{f2}$= 0.001, $t_{d3}$= 0.0001, $t_{f3}$= 0.0001, $\varepsilon_f$ = − 0.002, and the hybridization parameter $V$ = 0.0001 (negligible compared to $t_{d1}$).

Since there are several examples of complete orthogonal systems, viz, the Legendre polynomials over [-1,1], and Bessel function over [0,1], and so on, one may ask why show preference for $\{\sin(nx),\cos(nx)\}$ over $[-\pi, \pi]$. The reason is this is simpler to handle and replicates the model of a periodic crystal potential terminating at the surface where it undergoes a jump abruptly to the vacuum level. Now from the expectation value of the bulk Hamiltonian in (3), in the basis ($d^\dagger_{k,\uparrow}$ $d^\dagger_{k,\downarrow}$ $s^\dagger_{k,\uparrow}$ $s^\dagger_{k,\downarrow}$)$^T$, the Hamiltonian could be written in the matrix form as

$$\begin{pmatrix} \Gamma_{10} & 0 & -2Vbs_z & -2Vb(s_x - is_y) \\ 0 & \Gamma_{10} & -2Vb(s_x + is_y) & 2Vbs_z \\ -2Vbs_z & -2Vb(s_x - is_y) & \Gamma_{20} & 2Vbs_z \\ -2Vb(s_x + is_y) & 2Vbs_z & 0 & \Gamma_{20} \end{pmatrix},$$

$$\Gamma_{10} = -\mu - \xi - \varepsilon^d_k,$$
$$\Gamma_{20} = -\mu + \xi - b^2\varepsilon^f_k + \lambda. \tag{23}$$

In the basis ($d^\dagger_{k,\uparrow}$ $s^\dagger_{k,\uparrow}$ $d^\dagger_{k,\downarrow}$ $s^\dagger_{k,\downarrow}$)$^T$, the same bulk Hamiltonian will be given as

$$\begin{pmatrix} \Gamma_{10} & -2Vbs_z & 0 & -2Vb(s_x - is_y) \\ -2Vbs_z & \Gamma_{20} & -2Vb(s_x - is_y) & 0 \\ 0 & -2Vb(s_x + is_y) & \Gamma_{10} & 2Vbs_z \\ -2Vb(s_x + is_y) & 0 & 2Vbs_z & \Gamma_{20} \end{pmatrix} \tag{24}$$

Therefore, upon using (24), $\aleph_{mn}^{slab}(k_x, k_y)$ consistent with Eq. (22) is given by $H_s$ which is equal to

$$\begin{pmatrix} \Gamma_1 & -i\,c_{\zeta=+1} & 0 & 0 \\ i\,c_{\zeta=+1} & \Gamma_2 & 0 & 0 \\ 0 & 0 & \Gamma_1 & i\,c_{\zeta=-1} \\ 0 & 0 & -i\,c_{\zeta=-1} & \Gamma_2 \end{pmatrix} \tag{25}$$

where $c_{\zeta=+1} = \frac{16Vba}{3d}, c_{\zeta=-1} = \frac{96Vba}{7d}$, $K = (k_x, k_y)$, and in the long wavelength limit $\Gamma_1(K) = -\mu - \xi - 6t_{d1} + t_{d1}K^2$, and $\Gamma_2(K) = -\mu + \xi + b^2\varepsilon_f - 6b^2 t_{f1} + b^2 t_{f1} K^2 + \lambda$. It may be noted that, in view of (14), $\Gamma_1(K) \approx \Gamma_2(K)$.

The Dirac matrices in Dirac basis is given by $\gamma_j$ where $j = (0,1,2,3)$, and

$$\gamma_0 = \begin{pmatrix} I & 0 \\ 0 & -I \end{pmatrix}, \gamma_j = \begin{pmatrix} 0 & \sigma_j \\ -\sigma_j & 0 \end{pmatrix} \tag{26}$$

where $\sigma_j$ are Pauli matrices. A fifth related matrix is usually defined as $\gamma_5 \equiv i\gamma_0\gamma_1\gamma_2\gamma_3 = \begin{pmatrix} 0 & I \\ I & 0 \end{pmatrix}$. We will use below the Dirac-matrix Hamiltonian (DMH) method **[23]** to ascertain whether SmB$_6$ is a weak or strong TKI. It must be emphasized here that when a slab is considered, the top and bottom of the slab interface with vacuum, and hopping events between the surface states on the top and bottom opens a finite gap at the $\Gamma$ point ($k = 0$). If there is breaking of inversion symmetry (IS), then lifting the degeneracy of the surface state sub-bands with opposite spin angular momenta occurs. In this case, as we shall show, there is no IS breaking and therefore the two surface states overlap without spin flip. Off-diagonal mass-like terms must be included in the surface Hamiltonian $H_s$ in a suitable basis, say, $( d^\dagger_{k,\zeta} \quad s^\dagger_{k,\zeta} \quad d^\dagger_{k,\zeta} \quad s^\dagger_{k,\zeta} )^T$ to characterize the opening of finite gap at the $\Gamma$ point. A matrix of the form

$$\begin{pmatrix} 0 & 0 & M_0 & 0 \\ 0 & 0 & 0 & M_1 \\ M_0 & 0 & 0 & 0 \\ 0 & M_1 & 0 & 0 \end{pmatrix}$$

where $M_1$ (hopping parameter for f- electrons) $\ll M_0$ (hopping parameter for d electrons) perhaps will be suitable to take care of the overlap scenario. In what follows we, however, choose the simpler matrix, namely $\gamma_5 M_0$, hoping that the choice is not going to affect the aim stated above. The outcome of the inclusion of the mass-like term is that there would not be chiral liquid on the surface. There is, however, possibility of helical liquid provided we are able to see that the system is a strong TI. We shall show below the suitability of the matrix $\gamma_5 M_0$. As a first step of the DMH method, one needs to check the following: We consider non-spatial symmetries. These are symmetries which do not transform different lattice sites into each other. There are three different non-spatial symmetries that need to be considered: particle-hole symmetry (PHS) P, chiral symmetry (CS), and anti-unitary time-reversal symmetry $\Theta$(TRS) [23]. If a complex conjugation operator K is multiplied on the left side by a unitary matrix U, the resultant matrix UK is referred to as anti-unitary. It can be easily seen that for $4 \times 4$ matrices $\Theta = \Sigma_y K$ (this definition ensures $\Theta$ anti-unitary) , $P = \Sigma_x K$, and $C = $ diag(1, 1, $-1, -1$) where $\Sigma_j = \begin{pmatrix} \sigma_j & 0 \\ 0 & \sigma_j \end{pmatrix}$ and $\sigma_j$ are Pauli matrices. The properties of Pauli and four by four $\Sigma$ matrices are similar. The mass term $\gamma_5 M_0$ that leads to the opening of a gap at the band crossing must satisfy $[\Theta, \gamma_5 M_0] = 0$, $[P, \gamma_5 M_0] = 0$, and $\{C, \gamma_5 M_0\} = 0$ which we find they do. The second step is to check the anticommutativity of the matrix $\gamma_5 M_0$ with the surface Hamiltonian H$_s$ given by (25). If they do not anti-commute (do anti-commute), i.e. any sort of correlation does not

exist(exits) between $\gamma_5 M_0$ and $H_s$, then the band crossing cannot be gapped out (can be gapped out) which means the band crossing is topologically stable (unstable). We obtain here $\{H_s, \gamma_5 M_0\} \neq 0$, and therefore the topological stability of the band crossing is seemingly possible. The surface state Hamiltonian that we proceed with now is $H^{slab}(k_x, k_y) = H_s(k_x, k_y) + \gamma_5 M_0$. It is easy to see that the Hamiltonian $H^{slab}(k_x, k_y)$ commutes with the operator $\Theta$. Consequently, $H^{slab}(k_x, k_y)$ needs to satisfy the identity $H(-k) = \Theta H(k) \Theta^{-1}$. In view of $\Gamma_1(K) = \Gamma_2(K)$, it can be easily checked that $H^{slab}(-k_x, -k_y) = \Theta H^{slab}(k_x, k_y) \Theta^{-1}$. The identity implies that if energy band E(k), of a time-reversal(TR)symmetric system, satisfies E(k) =E(−k)) for a TR-invariant momentum (TRIM) pair ( k ,−k ) called Kramers pair, the relation −k +G = k (where G is a reciprocal lattice vector) is satisfied by such a pair due to the periodicity of the BZ. Also, one can gap out the helical edge states by introducing a Zeeman term that explicitly breaks the protecting time-reversal symmetry. As we will be showing below, we do obtain a term referred to as the pseudo-Zeeman term (PZT) in the single-particle spectrum which has different sign for opposite spins and no connection with momentum. Now, usually, the effects of an external magnetic field, B, perpendicular to the TI film are captured by two additional terms in the film Hamiltonian. The first term describes the orbital coupling to magnetic field through the minimal coupling $k_x \rightarrow k_x + (e/c)A_x$, where A = [−By, 0] is the vector potential in the Landau gauge. The second term describes the coupling of the spins to the magnetic field and is given by the Zeeman contribution. Thus, the PZT term in question does not act like a magnetic field here in real sense. The Hamiltonian $H^{slab}(k_x, k_y) = H_s(k_x, k_y) + \gamma_5 M_0$ is spin non-conserving, as $[H^{slab}(k_x, k_y), \Sigma_z] \neq 0$. The reason is the presence of the spin-orbit coupling. In fact, due to the spin-orbit coupling, the f-states are eigenstates of the total angular momentum J, and hence hybridize with conduction band states with the same symmetry. Furthermore, we find that

$$C^{-1} H^{slab}(k_x, k_y) C \neq H^{slab}(k_x, k_y). \qquad (27)$$

The absence of chiral symmetry ensures that the Hamiltonian $H^{slab}(k_x, k_y)$ could not be brought to an off-diagonal form by a unitary transformation. For example, in a Su-Schrieffer-Heeger[29] chain one can move the end states away from zero energy by breaking the chiral symmetry at the surface and/or in the bulk. We also obtain

$$P^{-1} H^{slab}(-k_x, -k_y) P \neq -H^{slab}(k_x, k_y), \qquad (28)$$

i.e. P-H Symmetry breaking. We wish to identify the effects of P-H symmetry breaking on the magneto-optical conductivity of the system in a sequel to this paper. Finally, the space inversion operator $\Pi$ (the eigenvalue of the $\Pi$ operator is ±1 as $\Pi^2 = 1$ ), acting on a state vector in position/momentum representation, brings about the transformations x→−x , σ→ σ, and k→−k where x stands for spatial coordinate, σ signifies the spin, and k stands for momentum. For $H^{slab}(k_x, k_y)$, it follows that $\Pi H^{slab}(k_x, k_y) \Pi^{-1} = H^{slab}(k_x, k_y)$, where $H^{slab}(-k_x, -k_y) = H^{slab}(k_x, k_y)$. Hence, the system preserves inversion symmetry (IS).

The surface state Hamiltonian that we proceed with now is $H^{slab}(k_x, k_y) = H_s(k_x, k_y) + \gamma_5 M_0$. This is TRS and IS invariant. But it breaks PHS and CS. We definitely require a version of surface state Hamiltonian better than $H^{slab}(k_x, k_y)$, say, under open boundary condition, by guessing an improved ansatz. The work is under way in this direction. The equation to find the eigenvalues of the Hamiltonian is a quartic. After lengthy algebra, this yields the following roots $E_{\alpha,\ surface}^{(\zeta)}(K) = \{E_1, E_2, E_3, E_4\}$ in view of the Ferrari's solution of a quartic equation:

$$E_{\alpha,\ surface}^{(\zeta)}(K) = \zeta\sqrt{\frac{z_0(K)}{2}} + \frac{(\Gamma_1(K)+\Gamma_2(K))}{2} + \alpha\left(b_0(K) - \left(\frac{z_0(K)}{2}\right) + \zeta c_0(K)\sqrt{\frac{2}{z_0(K)}}\right)^{\frac{1}{2}}, (29)$$

where $\zeta = \pm 1$ is the spin-index and $\alpha = \pm 1$ is the band-index. The first term $\sqrt{(z_0/2)}$ acts as an in-plane Zeeman term. The pseudo-Zeeman term of the spectrum (29) comes into being due the presence of the mass-like term $M_0$. Without this term, the spectrum reduces to a bi-quadratic rather than a quartic. The possible role of this term is that of the polarization-usherer. The other functions appearing in (29) are defined as follows:

$$E_{\alpha,\ surface}^{(\zeta=+1)}(K) = \{E_1(K), E_2(K)\} = \sqrt{\frac{z_0(K)}{2}} + \frac{(\Gamma_1(K)+\Gamma_2(K))}{2}$$
$$+ \alpha\sqrt{b_0(K) - \left(\frac{z_0(K)}{2}\right) + c_1^2},$$

$$E_{s,\ surface}^{(\zeta=-1)}(K) = \{E_3(K), E_4(K)\} = -\sqrt{\frac{z_0(K)}{2}} + \frac{(\Gamma_1(K)+\Gamma_2(K))}{2} +$$
$$+ \alpha\sqrt{b_0(K) - \left(\frac{z_0(K)}{2}\right) - c_1^2},$$

$$c_1^2 = c_0(K)\sqrt{\frac{2}{z_0(K)}}.$$

$$z_0(K) = \frac{2b_0(K)}{3} + \left(\frac{1}{2}\Delta^{\frac{1}{2}}(K) - A_0(K)\right)^{\frac{1}{3}} - \left(\frac{1}{2}\Delta^{\frac{1}{2}}(K) + A_0(K)\right)^{\frac{1}{3}},$$

$$A_0(K) = \left(\frac{b_0^3(K)}{27} - \frac{b_0(K)d_0(K)}{3} - c_0^2(K)\right), b_0(K) = \frac{3B^2(K) - 8C(K)}{16},$$

$$c_0(K) = \frac{-B^3(K) + 4B(K)C(K) - 8D}{32},$$

$$d_0(K) = \frac{-3B^4(K) + 256E(K) - 64B(K)D(K) + 16B^2(K)C(K)}{256},$$

$$\Delta(K) = \left(\frac{8}{729}b_0^6 + \frac{16d_0^2 b_0^2}{27} + 4c_0^4 - \frac{4d_0 b_0^4}{81} - \frac{8c_0^2 b_0^3}{27} + \frac{8c_0^2 b_0 d_0}{3} + \frac{4}{27}d_0^3\right),$$

$$B(K) = -2(\Gamma_1(K) + \Gamma_2(K)),$$
$$C(K) = [(\Gamma_1(K) + \Gamma_2(K))^2 + 2\Gamma_1(K)\Gamma_2(K) - \left(\frac{96Vba}{7d}\right)^2 - \left(\frac{16Vba}{3d}\right)^2 - 2M_0^2],$$

$$D(K) = \big(\Gamma_1(K) + \Gamma_2(K)\big)\big[-2\,\Gamma_1(K)\,\Gamma_2(K) + 2M_0^2 + \left(\tfrac{96Vba}{7d}\right)^2 +$$
$$\left(\tfrac{16Vba}{3d}\right)^2\big],$$

$$E(K) = M_0^4 + 2\,M_0^2\Gamma_1(K)\,\Gamma_2(K) + (\Gamma_1(K)\,\Gamma_2(K))^2 - \left(\left(\tfrac{96Vba}{7d}\right)^2 + \right.$$
$$\left.\left(\tfrac{16Vba}{3d}\right)^2\right)(\Gamma_1(K)\,\Gamma_2(K)) - M_0^2(\Gamma_1(K) + \Gamma_2(K))^2 + \left(\left(\tfrac{96Vba}{7d}\right)^2 \times \left(\tfrac{16Vba}{3d}\right)^2\right) - 2$$
$$M_0^2 \times \left(\left(\tfrac{96Vba}{7d}\right) \times \left(\tfrac{16Vba}{3d}\right)\right) \tag{30}$$

The surface states correspond to the eigenstates corresponding to the eigenvalues in (29). In an effort to examine the possibility of helical spin liquids, we have plotted the surface state energy spectrum given by Eq. (29) for dimensionless wavevector as a function of the dimensionless wave vector ($aK$) in Figure 4 for finite mass-like coupling and the various values of the chemical potential μ. Since the conduction bands are partially empty, the surface state will be metallic. In all figures, $t_{f1}$ is negative and therefore the figures correspond to insulating bulk. The important question is have we been able to show $SmB_6$ as a strong/ weak TI? The answer is not yet. For this purpose, we recall there is a key distinction between surface states in a conventional insulator and a topological insulator. The explanation is given below: We first recall what has been stated above. The TRS identity $H^{slab}(-k_x, -k_y) = \Theta H^{slab}(k_x, k_y)\Theta^{-1}$ satisfied here means that the energy bands come in (Kramers) pairs. The pairs are degenerate at the TR-invariant momentum (TRIM) where +**K** becomes equivalent to −**K** due to the periodicity of the BZ, i.e. **K** +**G** = −**K** where **G** is a reciprocal lattice vector. Now consider Figures 4(a) - (d). An inspection yields that, in the band structure displayed in 4(a), TRIM is **K** = (±1,0) /(0, ±1), in 4(b) also **K** = (±1,0) /(0, ±1), in 4(c) **K** = (±2,0) /(0, ±2), and in 4(d) **K** = (±3,0) /(0, ±3). The reason is that **K**'s satisfy the condition $\mathbf{K} + \mathbf{G} = -\mathbf{K}$. For example, **K** = (±1,0) and $\mathbf{G} = (\mp 2,0)$ added together will give $-\mathbf{K} = (\mp 1,0)$. In figure 4(c), the momentum **K**= (±2,0) or(0, ±2) obey **K**+ **G** = −**K** where **G** is the reciprocal lattice vector (∓4,0) or (0, ∓4). Similarly, in figure 4(d), the momentum **K**= ( ±3,0) or (0, 3±) obey **K**+ **G** = −**K** where **G** is (∓6,0) or (0,∓6). Let us now note that the Fermi energy $E_F$ inside the gap intersects these surface states (in the surface multi-band structures) in the same BZ either an even or odd number of times. If there are odd number of intersections at TRIMs, which guarantees the time reversal invariance, the surface state is topologically non-trivial (strong topological insulator), for disorder or correlations cannot remove pairs of such surface state crossings (SSC) by pushing the surface bands entirely above or below the Fermi energy $E_F$. This has been checked in Figure 4 by assigning different values to the chemical potential. We notice that the number of paired SSCs in each of the figures in Figure 4 is three(odd). However, amongst the three pairs of momenta (**K**,−**K**), only one in each of the figures satisfies the relation $-\mathbf{K} = \mathbf{K} + \mathbf{G}$. So, the number of TRIM involved in SSC is one(odd). When there are even number of pairs of surface-state crossings, the surface states are topologically trivial (weak TI or conventional insulator), for disorder or correlations can remove pairs of such crossings. The inescapable conclusion is that the system under consideration is a strong topological insulator. The material band structures are also characterized by Kane-Mele index $Z_2$= +1 ($v_0$ = 0) and $Z_2$= −1 ($v_0$ = 1). The former corresponds to weak TI, while the latter to strong TI.  The symmetry analysis and graphical representations lead to conclusion that the system considered here is a strong TI. The TKI surface, therefore, comprises of 'helical liquids'[30] in the slab geometry, which (helicity) is one of the most unique properties of a topologically protected surface. A recent report of a spin-signal on the surface by the inverse Edelstein effect [31] has confirmed the helical spin-structure.

## 4. Discussion and conclusion

We have started with periodic Anderson model (PAM) - a model for a generic TKI in the section 2 of the present communication. The model is quadratic in creation and annihilation operators without the on-site repulsion ($U_f$) between the f-electrons. The quadratic Hamiltonian makes the calculation of the relevant Matsubara propagators extremely simple. To investigate the model with $U_f$, the slave boson technique (SBT) [24] is employed in the mean-field theoretic(MFT) framework. The technique falls back on the conjecture that electrons can transmute into spinons and chargons (s-c). The s-c bound state needs to be fermionic to preserve the Fermi-Dirac statistics of the electrons. The two ways to have it on board are, (i) if the spinon is fermionic, the chargon should be bosonic (slave-boson) and, on the other hand, (ii) if the spinon is bosonic, the chargon should be fermionic (slave-fermion). Suppose, now the operators $b^†$ (light slave-boson creation operator) and $c^†$ (heavy slave-boson creation operator), respectively, creates empty bosonic impurity states and doubly-occupied states. Since the latter is prohibited for large on-site repulsion ($U_f >> t_{d1}$) between the f-electrons, the operators $c^†$ and c need not be taken into account when $U_f >> t_{d1}$. However, we need to take into account the remaining auxiliary operators ($s^†$, $b^†$) where the single site fermionic (bosonic) creation operator is denoted by $s^†$ ($b^†$). The operator $s^†$ ($b^†$) corresponds to the electron's spin(charge). The link between these auxiliary operators ($s^†$, $b^†$) to the physical f- electron operator is $f^†_\zeta = s^†_\zeta b$. In our SBT-MFT framework with $U_f >> t_{d1}$, we make the further assumption that slave-boson field (b) at each lattice site can be replaced by a c-number. The crucial anti-commutation relation $\{f_\zeta, f^†_{\zeta'}\} = \delta_{\zeta\zeta'}$ between the physical f- electron operators implies that the auxiliary operators will then satisfy the relation $\{s_\zeta, s^†_{\zeta'}\} = b^{-2} \delta_{\zeta\zeta'}$. Furthermore, the restriction $\sum_\zeta s^†_\zeta s_\zeta + b^† b = 1$ or, $\sum_\zeta \langle s^†_\zeta s_\zeta \rangle \cong 1 - b^2$ at a site needs to be imposed to take care of the conservation of auxiliary particle number. Thus, the complications associated with the large on-site repulsion ($U_f >> t_{d1}$) between the f-electrons is conveniently circumvented in the SBT by imposing a constraint to remove the double occupancy. The signature of the assumption "large on-site repulsion ($U_f >> t_{d1}$) between the f-electrons" is being carried over by the parameter λ and b which are present in $\Gamma_2$ (the defining equation could be found below (25)). The term $\Gamma_2$ is very significant for the surface state energy spectrum given by (29). Thus, the outcome noted above, viz. surface band structure showing odd number of crossing, is a consequence of the assumption. It will be interesting to re-investigate the problem with $U_f > t_{d1}$ (i.e. the on-site repulsion between the f-electrons is moderately strong) where the double occupancy is permissible. The slave-particle mean-field-theory is valid at very low values of temperature($T \to 0$). Moreover, we have neglected the dynamics of the boson field completely as the light slave-boson creation and destruction operators were replaced by a c-number'b'. This approximation has reduced the system to a non-interacting one with the bulk spectral gap dependent on the parameters (b, λ, ξ, V). Whereas the first two parameters take care of the strong correlation effect between the f-electrons, the hybridization parameter V is the harbinger of a topological dispensation.

In the expression of the momentum-dependent form-factor matrix Γ, we have taken into account only the lowest-order cubic harmonics[1,15,16]. In fact, our intention is to use more complicated odd-parity expressions for Γ and re-investigate the present problem bringing about some improvement in the surface Hamiltonian. This is expected to facilitate more refined analysis of odd/even number crossing issue. We summarize below our progress for the benefit of reader. Let us consider the momentum-dependent form-factor matrix involved in the third term in Eq.(1). This term actually is given by

$$\sum_{k,m_j\zeta=\uparrow,\downarrow}\{\langle k,\zeta=\uparrow,\downarrow|V_0|\Gamma^f:R=0,m_j\rangle d^\dagger_{k\zeta}f_{m_j}+\text{H.C.}\}$$

where the fermionic operator $d^\dagger_{k\zeta}$ creates a Bloch state with k and $\zeta \in \{\uparrow,\downarrow\}$ being momentum and physical spin quantum numbers, respectively. The fermionic operator $f^\dagger_{m_j}$ creates a localized Kramers doublet state $|\Gamma^f:R=0,m_j\rangle$ associated with the crystal field effect and spin-orbit coupling. In the case of $SmB_6$ the lowest-lying Kramers doublet is $|\Gamma_8^{f(2)}\rangle = |m_j = \pm\frac{1}{2}\rangle$. So, $m_j \in \{\uparrow,\downarrow\}$ is a pseudospin quantum number, corresponding to the two possible values of the projection of total angular momentum in the lowest-lying Kramers doublet. We assume that the hybridization operator has a spherical symmetry in the vicinity of the heavy atom. Also, we expand the Bloch wave of s = 1/2 fermion in terms of spherical harmonics and partial waves. Besides, the lowest-lying Kramers doublet state can be can be decomposed into radial and angular part. We further assume a constant hybridization amplitude (CHA). One can now rewrite the effective hybridization matrix elements by using these assumptions as a product of CHA, Clebsch-Gordan coefficient for s = ½ and the spherical harmonics. One can then calculate the hybridization matrix for two interesting cases, viz. (i) l = 1, s = 1/2, j = 1/2 and $m_j = \pm 1\,2$ (ii) l = 3, s = 1/2, j = 5/2 and $m_j = \pm 1/2$. In the former case, we find the form factor is equal to a product of $V_0$, an odd spin, and odd momentum operator. Therefore, it is time-reversal invariant. In the latter case (which corresponds to $SmB_6$), it is of form $\{V_0 k^3 \det\}$ where 'det' is determinant involving spherical harmonics. All the matrix elements of the Hamiltonian (25) will now have to be correct up to $O(k^3)$ including the hopping terms for the consistency sake. One then expects to find better version of the surface state Hamiltonian to investigate the odd/even number of crossing issue. Furthermore, we shall have a more accurate platform of investigating theoretically the mysterious behavior of bulk insulator $SmB_6$ displaying metal-like Fermi surface even though there is no long-range transport of charge.

The effect of Bychkov − Rashba term on the present system is quite interesting. In order to show why is it so, we proceed with new surface state Hamiltonian $H^{slab}(k_x,k_y) = H_s(k_x,k_y) + $ Bychkov − Rashba tervm, where the two by two matrix equivalent of the latter is $A_0 = [\lambda_R(k_x\sigma_y - k_y\sigma_x)]$ and $\lambda_R$ is the strength of the interaction. It is imperative to assume that there must be metals such as Au(111), on the surface in close proximity or deposition of particles with considerably high Rashba spin-orbit (RSO) interaction strength on the surface of material. Assuming that the Rashba interaction is between f-electrons only, we write the surface state Hamiltonian, in the basis $(d^\dagger_{k,\uparrow}\ s^\dagger_{k,\uparrow}\ s^\dagger_{k,\downarrow}\ d^\dagger_{k,\downarrow})^T$, as

$$\aleph_{mn}^{slab}(k_x,k_y) =$$

$$\begin{pmatrix} \Gamma_1 & -iA_1 & 0 & 0 \\ iA_1 & \Gamma_1 & A^*_3 & 0 \\ 0 & A_3 & \Gamma_1 & -iA_2 \\ 0 & 0 & iA_2 & \Gamma_1 \end{pmatrix}, \quad (31)$$

where $A_3(K) = \lambda_R(-ik_x - k_y)$, $A_1 = c_{\zeta=+1}$, and $A_2 = c_{\zeta=-1}$. The equation to find the eigenvalues of the Hamiltonian is a bi-quadratic. This yields the following roots

$$E^{(\zeta)}_{\alpha,\,surface}(K) = \Gamma_1(K) + \left(\frac{\alpha}{2}\right)\left(\left(A^2_1 + A^2_2 + |A_3(K)|^2\right) + \zeta c_0(K)\right)^{1/2}, \quad (32)$$

where $c_0(K) = \left(\left(A^2_1 + A^2_2 + |A_3(K)|^2\right)^2 - 4A^2_1 A^2_2\right)^{1/2}$, $\alpha = \pm 1$ is the band-index, and $\zeta = \pm 1$ is the spin-index. Plots of surface state single-particle excitation spectrum given by Eq.(32) versus dimensionless wave vector is shown in Figure 5. The parameter values are b = 0.95, $t_{d1}$ = = 1, $t_{f1}$ = -0.35, $\varepsilon_f$ = -0.06, µ= -0.50, V = 0.10, $V_p = \left(\frac{Vab}{d}\right) = 0.05$, A$_1$= 5.3333.*Vp, A$_2$ = 13.7143.*Vp, and $\lambda_R = 0.81$. Since all four bands are partially empty, the surface state will be metallic. The hopping integral $t_{f1}$ is negative and therefore the system corresponds to insulating bulk. In Figure 5, the two bands

$$\Gamma_1(K) + \left(\frac{1}{2}\right)\left(\left(A^2_1 + A^2_2 + |A_3(K)|^2\right) - c_0(K)\right)^{1/2}, \quad (33)$$

and

$$\Gamma_1(K) - \left(\frac{1}{2}\right)\left(\left(A^2_1 + A^2_2 + |A_3(K)|^2\right) + c_0(K)\right)^{1/2}, \quad (34)$$

appear to be degenerate. They are, in fact, not so as that would demand fulfilment of the inadmissible condition $c_0(K) = 0$. Actually, they are too close to appear as resolved in Figure 5(a). As we have noted, if there are an odd (even) number of TRIM related surface state crossings (SSC), the surface states are topologically non-trivial (trivial). We find that, with greater resolution, there are no TRIM involved SSCs (the apparent near-degeneracy of (33) and (34) may mislead us to believe that there are even number of TRIM linked SSCs) in Figure 5 (b). This means disorder or correlations can remove pairs of such crossings by pushing the surface bands entirely above or below the Fermi energy E$_F$. Seemingly, the broken IS due to the Bychkov − Rashba term, ushers in the change from the strong topological insulator to a conventional one. A detailed investigation is required before we announce the final verdict.

In conclusion, we are aware of the fact that the investigation presented does not take into account of the specific electronic structure of the real TKI system, such as SmB$_6$. This makes the work less useful. In a sequel to this paper this issue will be addressed. Moreover, we recall that, in a slab geometry, the top and bottom of the slab interface with vacuum and hopping events between the surface states on the bottom and top opens a finite gap at the $\Gamma$ point (k = 0). Off-diagonal mass-like terms must be included in the surface Hamiltonian $H_s$, in a suitable basis, to characterize the opening of this finite gap. The point we wish to make is that while, in hopping, a particle has to have energy greater than or equal to that of the height of the potential barrier (in between atoms at the top and bottom) in order to cross the barrier, in tunneling the particle can cross the barrier even with energy less than the height of the barrier. Besides, the former is classical whereas the latter, which depends on the width of the barrier, is a quantum-mechanical phenomenon. Thus, our assumption of the 'occurrence of the hopping only' needs to be supplemented with an account of the tunneling. In fact, a full-proof description must account for all processes that make an electron cross from one atom to another.

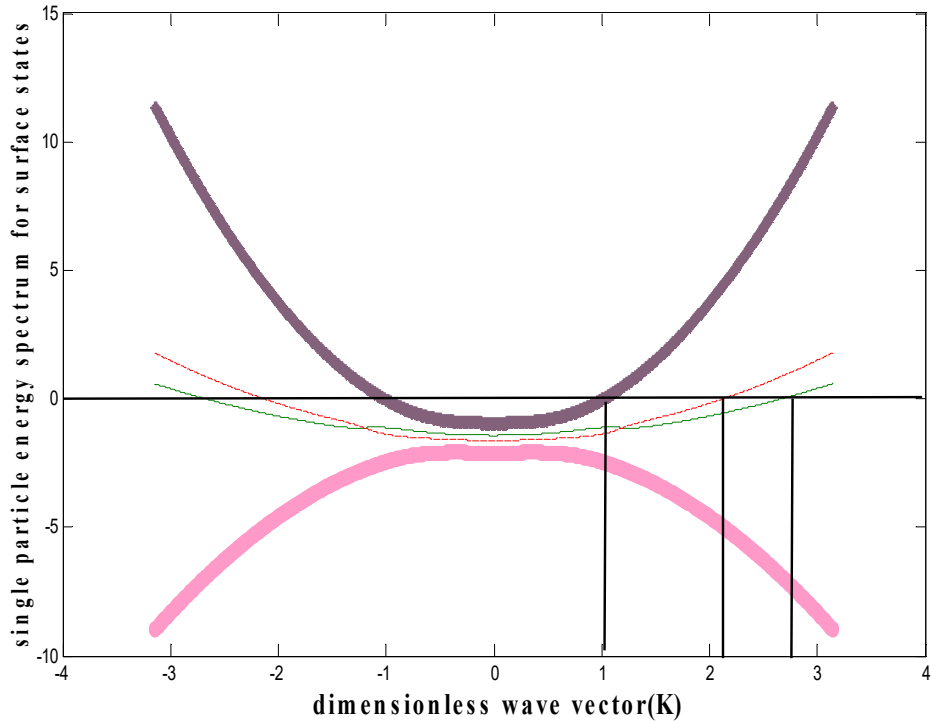

(a)

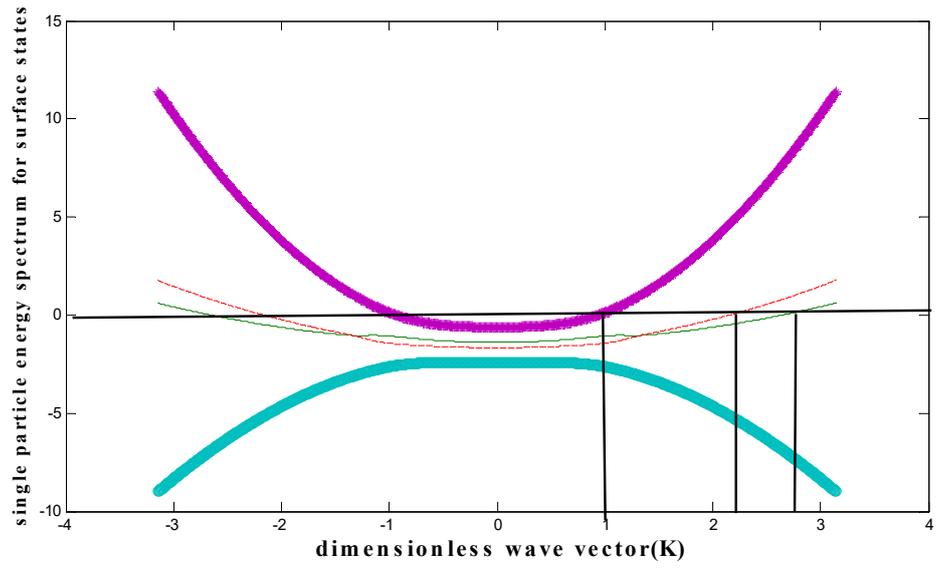

**(b)**

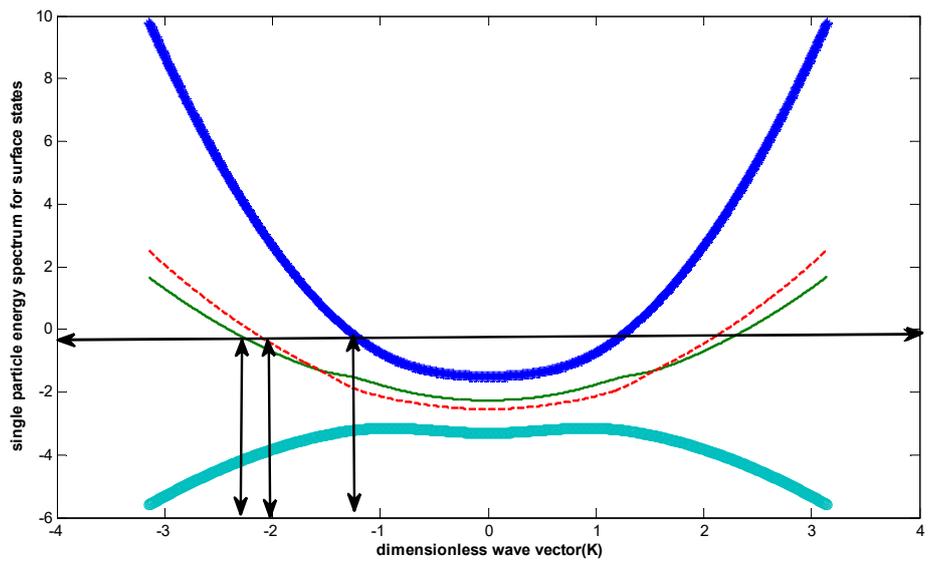

**(c)**

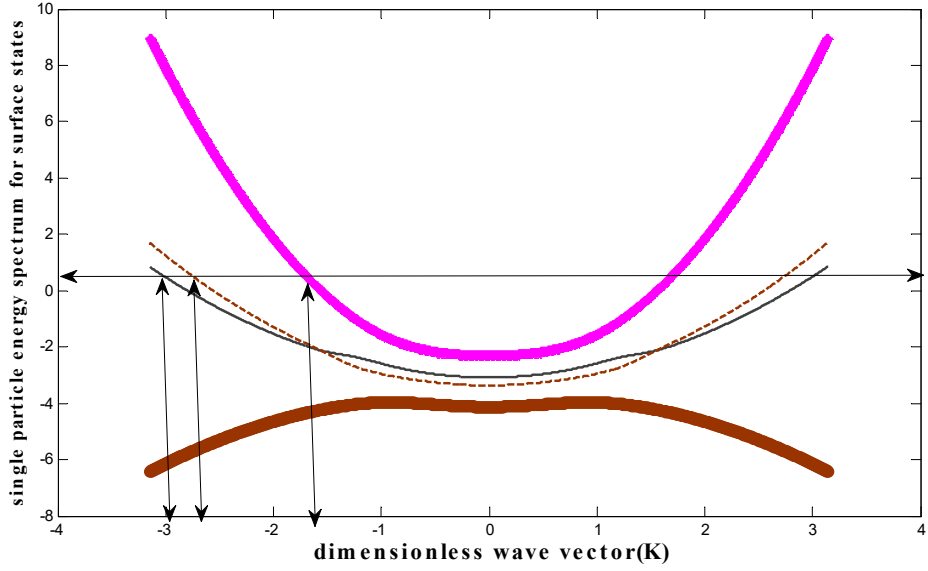

**(d)**

**Figure 4.** The plots of surface state single-particle excitation spectrum given by Eq.(29) versus dimensionless wave vector for the various values of chemical potential μ and the mass-like term $M_0$ : **(a)** μ= 0.00, $t_{f1} = -0.5$ and $M_0 = 0.02$, **(b)** μ= 0.00, $t_{f1} = -0.50$ and $M_0 = 0.50$, **(c)** μ= -0.32, $t_{f1} = -0.1$, and $M_0 = 0.50$, and **(d)** μ= -0.50, $t_{f1} = -0.10$ and $M_0 = 0.50$. The other parameters are b = 0.95, $t_{d1}= 1$, $\varepsilon_f = -0.05$, $V = 0.10$, $V_p = \left(\frac{Vab}{d}\right) = 0.05$. Since the conduction bands are partially empty, the surface state will be metallic in all the cases. The hopping integral $t_{f1}$ is negative and therefore the system corresponds to insulating bulk. The number of TRIM involved pair crossing in each of the figures is one(odd).

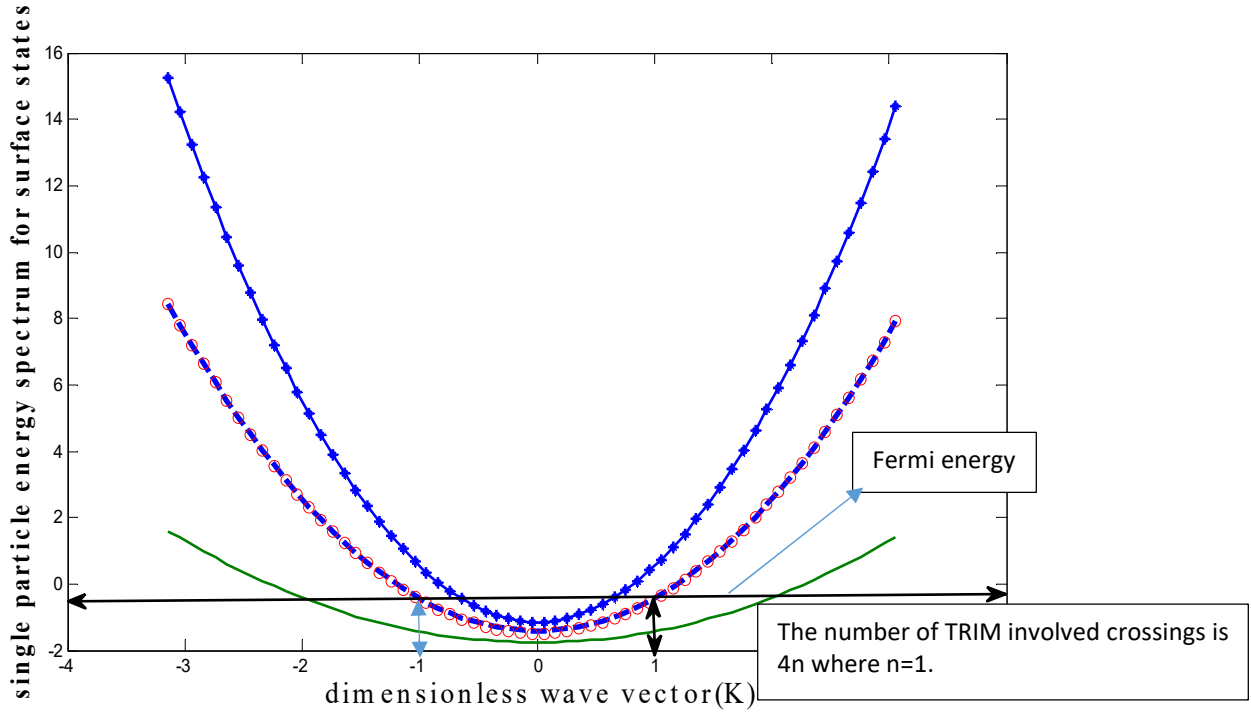

**(a)**

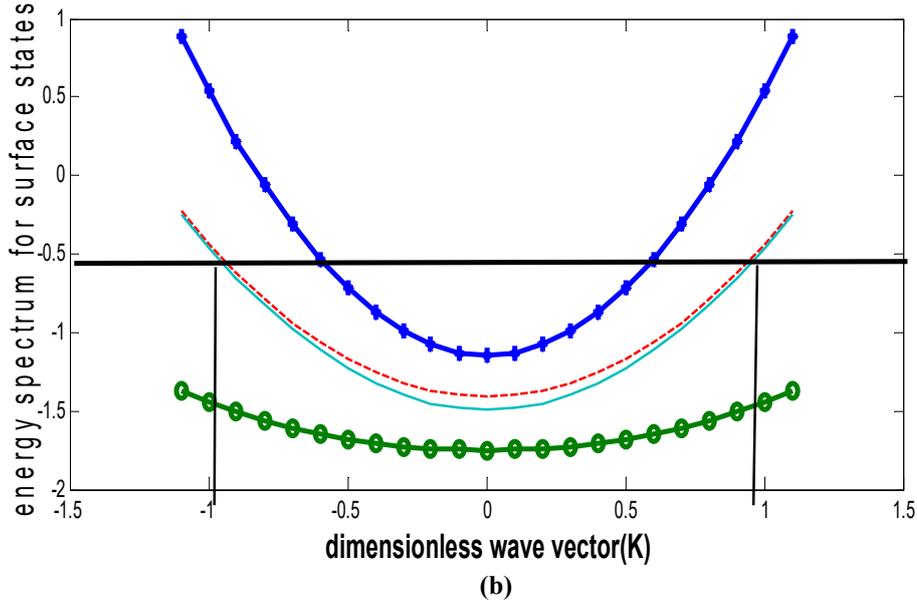

**Figure 5.** Plots of surface state single-particle excitation spectrum given by Eq.(32) versus dimensionless wave vector. The parameter values are b = 0.95, $t_{d1}$ = = 1, $t_{f1}$= -0.35, $\varepsilon_f$ = -0.06, μ= -0.50, V = 0.10, $V_p = \left(\frac{Vab}{d}\right)$ = 0.05, $A_1$= 5.3333.*Vp, $A_2$ = 13.7143.*Vp, and $\lambda_R$ = 0.81. Since all four bands are partially empty, the surface state will be metallic. The hopping integral $t_{f1}$ is negative and therefore the system corresponds to insulating bulk. (a) This figure apparently shows even number of TRIM involved surface state crossings (SSC). (b) This figure with greater resolution show that only non-TRIM momenta are involved in the SSC.